Title:

# Time-dependent Clearance of Cyclosporine in Adult Renal Transplant Recipients: A Population Pharmacokinetic Perspective

Junjun Mao[1], Xiaoyan Qiu[1*], Weiwei Qin[1*], Luyang Xu[1], Ming Zhang[2], Mingkang Zhong[1]

[1] Department of Pharmacy, Huashan Hospital, Fudan University, 12 Middle Urumqi Road, Shanghai 200040, China

[2] Department of Nephrology, Huashan Hospital, Fudan University, 12 Middle Urumqi Road, Shanghai 200040, China

**\*Corresponding authors:**

Xiaoyan Qiu, PhD

Professor of Clinical Pharmacy

Department of Pharmacy, Huashan Hospital, Fudan University

12 Middle Urumqi Road, Shanghai 200040, China

Tel.: (8621) 5288 8712; Fax: (8621) 62486927

Email: xyqiu@fudan.edu.cn

Weiwei Qin, PhD

Associate Professor of Clinical Pharmacy

Department of Pharmacy, Huashan Hospital, Fudan University

12 Middle Urumqi Road, Shanghai 200040, China

Tel.: (8621) 5288 8712; Fax: (8621) 62486927

Email: wwqin@fudan.edu.cn

**Submitting author:**

Junjun Mao

Email: jmao12@fudan.edu.cn

**Principal Investigator statement:**

The authors confirm that the Principal Investigator for this paper is Ming Zhang and that he had direct clinical responsibility for patients.

**Running head:**

Time-dependent clearance of CsA

**Word Counts:**

Abstract: 250; Main text (Introduction, Methods, Results and Discussion): 3993

**Number of tables and figures:** 3 and 4

**Number of supplemental texts, tables and figures:** 3, 4 and 1

**Number of references:** 54

## Abstract

**Aim**

The pharmacokinetic (PK) properties of cyclosporine (CsA) in renal transplant recipients are patient- and time-dependent. Knowledge of this time-related variability is necessary to maintain or achieve CsA target exposure. Here, we aimed to identify factors explaining variabilities in CsA PK properties and characterise time-dependent clearance (CL/*F*) by performing a comprehensive analysis of CsA PK factors using population PK (popPK) modelling of long-term follow-up data from our institution.

**Methods**

In total, 3,674 whole-blood CsA concentrations from 183 patients who underwent initial renal transplantation were analysed using nonlinear mixed-effects modelling. The effects of potential covariates were selected according to a previous report and well-accepted theoretical mechanisms. Model-informed individualised therapeutic regimens were also conducted.

**Results**

A two-compartment model adequately described the data and the estimated mean CsA CL/*F* was 32.6 L $h^{-1}$ (5%). Allometrically scaled body size, haematocrit (HCT) level, CGC haplotype carrier status, and postoperative time may contribute to CsA PK variability. The CsA bioavailability in patients receiving a prednisolone dose (PD) of 80 mg was 20.6% lower than that in patients receiving 20 mg. A significant decrease (52.6%) in CL/*F* was observed as the HCT increased from 10.5% to 60.5%. The CL/*F* of the non-CGC haplotype carrier was 14.4% lower than that of the CGC haplotype carrier at 3 months post operation. CsA dose adjustments should be considered in different postoperative periods.

**Conclusions**

By monitoring body size, HCT, PD, and CGC haplotype, changes in CsA CL/*F* over time could be predicted. Such information could be used to optimise CsA therapy.

**What is already known about this subject:**

- The pharmacokinetic properties of CsA in renal transplant recipients are patient- and time-dependent.

- The bioavailability of CsA increases rapidly in the immediate postoperative period and then decreases gradually to reach the initial value approximately 1 - 1.5 years after transplantation.

- Saturable presystemic metabolism and the effects of *NR1I2* polymorphism contribute to this tendency.

**What this study adds:**

- The first popPK study of long-term changes (> 16 years) in CsA CL/*F* with POD in adult renal transplant patients was conducted.

- Tapering of PD during the initial stage of transplantation may contribute to changes in bioavailability.

- By monitoring body size, HCT, PD, and *ABCB1* CGC haplotype, changes in CsA CL/*F* over time could be predicted.

# 1 Introduction

Cyclosporine (CsA), a potent calcineurin inhibitor, is commonly used to prevent allograft rejection after renal transplantation [1, 2]. With the introduction of CsA combination therapy, the survival rates of transplant patients, particularly short-term outcomes, have greatly improved [3]. However, prolonged use of CsA leads to substantial toxicity, which may reduce long-term renal allograft survival as well as increase the risk of cardiovascular morbidity and mortality [4-6]. Most adverse events and rejection rates may correlate with the concentration of CsA applied.

CsA exhibits unique pharmacokinetics (PK), including low bioavailability (approximately 25%; range: 10 - 89%) owing to its poor aqueous solubility and low transmembrane permeability in the intestine mediated by P-glycoprotein (P-gp) [1, 7]. In addition, CsA binds extensively to erythrocytes [8], is predominantly metabolised by cytochrome P450 (CYP) 3A isoenzymes [9], and is then eliminated in the bile [10]. CsA has a narrow therapeutic index and large inter- and intra-individual PK variability. Therefore, it is essential to conduct routine therapeutic drug monitoring (TDM) to optimise the CsA dosage regimen and minimise adverse effects [11].

Population PK (popPK), which is a superior approach to classical PK analysis, can be used to obtain population standard values and identify covariates with sparse sampling during TDM [12-14]. Currently, numerous popPK models have been developed to quantitatively describe the PK characteristics of CsA [15]. Several clinical factors, including body weight, postoperative time, and haematocrit (HCT), have been identified to explain the PK variability of CsA [16].

The PK parameters of CsA in renal transplant recipients are patient- and time-dependent. The dose required to achieve targeted whole-blood concentrations of CsA varies considerably among patients and according to the time after transplantation. Two distinct challenges exist for CsA dosing individualisation in transplant recipients, i.e., predicting the initial dose of CsA without any known concentration-time data in a particular patient and adjusting the dose over time after transplantation. To maintain or achieve CsA target exposure, knowledge of time-dependent PK characteristics is necessary.

The postoperative day (POD) is a combined reflection of various time-related factors, such as the recovery of gastrointestinal function [17] and tapering of the doses of co-administered steroids [18]. During the early stages after transplantation, gastrointestinal function is abnormal, which decreases the bioavailability of CsA. With the recovery of gastrointestinal function during therapy, CsA absorption improves, thereby decreasing the apparent clearance (CL/*F*) [19]. However, tapering the steroid dosage during immunosuppressive therapy may reduce the expression of CYP3A and P-gp, leading to increased absorption and accelerated metabolism of CsA [20]. Moreover, an increase in the HCT leads to increased binding of CsA to erythrocytes, resulting in elevated concentrations and decreased CL/*F* [21-23]. In addition to the abovementioned considerations, other unidentified confounding factors may contribute to the complexity of the impact of POD. Therefore, a clearer understanding of POD as a covariate is required for dose optimisation.

Although various studies have been conducted to describe the tendency of CsA CL/*F* over POD [21, 24], most studies have had a short follow-up (less than 1 year), resulting in limited information regarding long-term time-related variability in CsA PK [15, 16]. Only one study conducted by Fanta *et al*. reported time-dependent CsA PK properties in a long-term follow-up (> 16 years) of paediatric patients [25]. They reported that the bioavailability of CsA increases rapidly in the immediate postoperative period and then decreases gradually to reach the initial value approximately 1 - 1.5 years after transplantation. Saturable presystemic metabolism and the effects of *NR1I2* polymorphism were found to contribute to this tendency. However, no such research has been reported in adults.

In this study, we aimed to identify factors explaining variability in CsA PK and characterise the time-dependent CL/*F* of CsA in a comprehensive analysis of the effects of demographic, clinical, and genetic factors on CsA PK using popPK modelling of long-term follow-up data available from our institution. Model-informed individualised therapeutic regimens were also evaluated.

## 2 Materials and Methods

### 2.1 Patients and data collection

Data were collected from 183 adults (122 men and 61 women) who underwent renal transplantation at Huashan Hospital. The inclusion criteria were as follows: (1) allograft renal transplantation for the first time, (2) age greater than 18 years, and (3) CsA-based triple immunosuppressive regimen used. The exclusion criteria were as follows: (1) administration of the conventional oral formulation of CsA, (2) undergoing dialysis treatment, and (3) required covariate data missing. Demographic and pathophysiological data were obtained during routine clinical visits from July 2003 to December 2016.

In total, 3,674 whole-blood CsA concentrations were available for model development. 3,326 sparse-PK samples of CsA predose concentrations ($C_0$) and 2-h postdose concentrations ($C_2$) were retrospectively collected from follow-up TDM. Additionally, 24 concentration-time profiles (259 full-PK samples) were evaluated during the 12 h after the morning dose of CsA. Whole-blood samples were mainly obtained before dosing and 0.5, 1, 1.5, 2, 3, 4, 6, 8, 10, and 12 h after dosing. In addition, 89 other time points were included (1, 2, 3, 4, and 6 h after dosing). All samples were stored at -20 ℃ until analysis.

The study protocols were approved by the Ethics Committee of Huashan Hospital and conducted in accordance with the Declaration of Helsinki. All patients provided written informed consent before enrolment in the study.

**2.2 Immunosuppressive therapy**

All patients were administered combined immunosuppressive therapy, including a CsA microemulsion (Neoral; Novartis Pharma Schweiz AG, Emberbach, Germany), mycophenolate mofetil (MMF; CellCept; Roche Pharma Ltd., Shanghai, China) and corticosteroids. The initial dose of CsA was 5 mg kg$^{-1}$ d$^{-1}$, administered as two doses under fasting conditions immediately after surgery. Subsequent doses were empirically adjusted to achieve target concentrations based on local guidelines (Text S1) [26]. MMF (0.5 - 3 g d$^{-1}$) was administered according to body size and POD. This schedule was followed with oral prednisolone (80 mg d$^{-1}$), and the dosage was gradually decreased by 10 mg d$^{-1}$ until reaching 20 mg d$^{-1}$ after 10 d. The dosage was further tapered to 15, 10, and 5 mg d$^{-1}$ by months 1, 3, and 6, respectively.

**2.3 Determination of CsA concentration**

Whole-blood samples were collected from July 2003 to April 2011 and analysed using a well-validated fluorescence polarisation immunoassay (FPIA) on an AxSYM Abbott diagnostic system (Abbott Diagnostics, Chicago, IL, USA). Samples collected from May 2011 to December 2016 were analysed using a chemiluminescent microparticle immunoassay (CMIA) on an Architect I2000 system (Abbott Diagnostics).

The following formula (Eq. 1) [27] was used to convert the CMIA measured $C_0$ before modelling, owing to the systematic biases and cross-reactivity of metabolites between the methods.

$$\text{AxSYM} = 0.87 \times \text{CMIA} + 25.84 \qquad (1)\ [27]$$

AxSYM represents the FPIA performed using an AxSYM Analyzer, whereas CMIA was performed on an Architect system, as described above.

For AxSYM, the limit of detection (LOD) was 21.8 ng mL$^{-1}$ and the calibration range was 40 - 800 ng mL$^{-1}$; for CMIA, the LOD was 25 ng mL$^{-1}$ and the calibration range was 30 - 1500 ng mL$^{-1}$.

**2.4 Genotyping and haplotype analysis**

Five single-nucleotide polymorphisms (SNPs), i.e., *CYP3A4\*1G, CYP3A5\*3,* and *ABCB1* C1236T, G2677T/A, and C3435T, were genotyped by an independent external contractor (GeneCore BioTechnologies Co., Ltd., Shanghai, China) using a DNA sequencing apparatus (Applied Biosystems 3730; Thermo Fisher Scientific, Waltham, MA, USA) [28]. Deviations from the Hardy-Weinberg equilibrium were examined using Pearson's $\chi^2$-test. Linkage disequilibrium (LD) between different pairs of *ABCB1* SNPs was determined using the absolute standardised LD coefficient. Further details are available in Text S2.

**2.5 PopPK model development**

PopPK modelling was developed using nonlinear mixed-effects modelling software (NONMEM version 7.4; ICON Development Solutions, Ellicott City, MD, USA), with Pirana 2.9 as an interface for Perl Speaks NONMEM (PsN; version 4.9.0) [29]. Graphical analyses were processed using R software (version 3.5.0; http://www.r-project.org/). The first-order conditional estimation method, including η-ε interactions (FOCE-I), was employed throughout the method-building procedure [30].

**2.5.1 Base model**

Based on visual inspection of the data and a review of the literature, a two-compartment model with first-order absorption and lag time was used to describe CsA PK [16]. The estimated parameters included CL/*F*, central volume of distribution ($V_c$/*F*), inter-compartmental clearance (Q/*F*), apparent peripheral volume of distribution ($V_p$/*F*), absorption rate constant ($K_a$), absorption lag time ($T_{lag}$), and bioavailability (*F*) relative to a population standard value defined as 1.

Between-subject variability (BSV) was estimated for all parameters, except $T_{lag}$, and was assumed to be log-normally distributed. Proportional and combined proportional as well as additive structures were tested to describe the residual unexplained variability.

**2.5.2 Covariate model**

Demographic and pathophysiological data, as well as concomitant medications (Table 1), were evaluated as covariates. Data extracted from medical records, including body size, HCT, PODs, and prednisolone dose (PD), were evaluated as possible covariates of CsA PK. These covariates were selected according to a previous report and their clinical relevance [16]. The correlation between preselected covariates and the changes of covariates with POD were investigated graphically.

As the most frequently identified covariate in the final models, the change in CL/*F* and volume of distribution as a function of body weight was empirically described allometrically [31, 32]. Body size was based on fat-free mass (FFM) predicted from total body weight, height, and sex (Text S3) [33].

Considering the hypothesis that tapering PD may significantly influence bioavailability [34, 35], the relationship between PD and *F* was described using an $E_{max}$ model (Eq. 2), as follows:

$$F = F_{20} \times \left(1 + \frac{F_{\max} \times (PD-20)}{ED_{50} + (PD-20)}\right) \qquad (2)$$

$F_{20}$ is the bioavailability of CsA for a PD of 20 mg, assumed to be 1. Parameters $F_{max}$ and $ED_{50}$ are the maximal increase in $F$ and the dose above 20 mg, which corresponds to half the $F_{max}$, respectively.

As for substrates of CYP3A and P-gp, we also comprehensively evaluated the influence of genotype, including single gene sites, combined genotypes, and haplotypes of the drug-metabolising enzymes *CYP3A5\*3* and *CYP3A4\*1G* and the multidrug resistance transporter *ABCB1* with C1236T, G2677T/A, and C3435T polymorphisms.

Potential covariates were screened using a stepwise approach. The influence of other continuous covariates was explored using a linear, exponential, and power function model. Categorical variables, such as SNPs and concomitant medications, were investigated by estimating the fractional change in one group compared with that in the other groups.

### 2.5.3 Model selection criteria

Visual model fit was evaluated using standard goodness-of-fit (GOF) criteria, reductions in the objective function value (OFV), and acceptable precision on estimates [30]. A covariate was considered significant if its inclusion decreased the OFV by more than 3.84 ($\chi^2$-test, $p < 0.05$, $df = 1$) and if backward elimination of the covariate increased the OFV by more than 10.83 ($\chi^2$-test, $p < 0.001$, $df = 1$). Moreover, a clear pharmacological or biological basis was also considered as covariates were added. During the model development process, the condition numbers were calculated, and no more than 1,000 were kept to avoid over-parameterisation [36].

### 2.6 Model evaluation

Fifty-six patients from the evaluation group were used to examine the predictability of the final model. The adequacy of the model was assessed using GOF plots and prediction-corrected visual predictive checks (pcVPCs) [37], and a nonparametric bootstrap was employed to assess the robustness of the model [38]. The dataset was simulated 2,000 times for the pcVPCs. The 95% confidence intervals for the median, and 5[th] and 95[th] percentiles of the simulations at different time points were calculated and graphically compared with the observations, using automatic binning.

For the nonparametric bootstrap procedure, 500 bootstrap datasets were generated by random sampling with replacement using Perl modules [39]. The median and the 2.5[th] to 97.5[th] percentiles of the parameters after bootstrap runs with successful convergence were compared with the final model parameter estimates.

### 2.7 Dosing regimen optimisation

Monte Carlo simulations were conducted using parameter estimates from the final model to determine the optimal starting dosing regimen and achieve the target concentration during different postoperative periods. The CsA dose was simulated from 50 mg q12h to 300 mg q12h for a standard-sized subject (FFM 50 kg) with different covariate levels (Table S1). Time-concentration profiles were simulated based on 1,000 hypothetical individuals, and the steady-state $C_0/C_2$ value for each simulated subject was calculated. The median and the $25^{th}$ to $75^{th}$ percentiles of a steady-state $C_0/C_2$ value were calculated to select the optimal dosing regimen.

## 3 Results

### 3.1 Patients

Demographic characteristics and clinical data of the study population are presented in Table 1. In total, 3,674 CsA whole-blood measurements were available from 183 renal transplant recipients. A median of 15 CsA observations was obtained for each patient (range, 2 - 50). A description of the sampling points is provided in Table S2. Sampling occasions varied from day 1 to day 5998 (> 16 years) after transplantation and CsA doses ranged from 25 to 300 mg twice daily.

All allele frequencies of *CYP3A4\*1G*, *CYP3A5\*3*, and *ABCB1* genetic polymorphisms were in Hardy-Weinberg equilibrium (Table 2). D' values between *ABCB1* G2677T/A or C1236T and C3435T were 0.69 and 0.75, respectively, whereas that between *ABCB1* G2677T/A and C1236T was 0.60. These results indicate that *ABCB1* C1236T and G2677T/A were in LD with C3435T. Haplotype frequencies of *ABCB1* C1236T-G2677T/A-C3435T were calculated using SHEsis [40]. Only haplotypes with frequencies and patient proportions of more than 8% were analysed (Table S3).

Patients were divided into two groups: data from 127 patients were used for model development, and data from 56 patients were used for model evaluation. Sixteen patients without genetic information were included in the evaluation dataset. No concentrations below the lower quantification limit were included in the analysis.

### 3.2 PopPK model development

#### 3.2.1 Base model

A two-compartment model with first-order absorption and lag time was selected as the base model to describe CsA PK. The exponential model provided the best results for the residual variability. The BSV of the mean CL/*F* in the base model was 24.2%, with a relative standard error of 9.6%. The parameter estimates and associated precisions are listed in Table 3.

### 3.2.2 Covariate model

Before the stepwise process, the correlation between PK parameters and potential covariates was investigated graphically. The CsA CL/*F* tended to decrease with POD after renal transplantation, then increased and reached stability (Figure 1), consistent with the results from Fanta *et al*. [25]. Therefore, POD was incorporated into the base model exponentially; the GOF improved, and the OFV decreased by -54.1 (Table 3). The GOF plots are presented in Figure 2.

Although the base model with POD could partly describe the changes in CsA CL/*F* over time, a model in which covariates describe this time dependency is more useful for TDM. First, the changes in covariates and POD were examined by graphical inspection to identify potential covariates (Figure S1). Then, the identified covariates were incorporated into the base model to test the effects on CsA PK properties.

The influence of patient body size on CsA disposition was best described by allometric scaling based on FFM rather than total body weight for all PK disposition parameters, with the OFV reduced by -11.2 ($P < 0.001$). Because the HCT was low early during the postoperative period and increased during the first months after transplantation, the effect of the HCT on CsA clearance was investigated. The OFV substantially declined by -41.4 ($P < 0.001$) when the HCT was included, indicating a significant model improvement. A decrease (52.6%) in CL/*F* was observed as the HCT increased from 10.5% to 60.5%.

As the PD was tapered dramatically during the initial stage of transplantation, the effects of prednisolone on *F* were added to the model as a high PD may influence CsA absorption nonlinearly. Therefore, a sigmoid $E_{max}$ model describing the effect of PD on *F* (Eq. 2) reduced the OFV by -24.4 ($P < 0.001$). $F_{max}$ and $ED_{50}$ were estimated as 23.8 mg and -0.288, respectively. The relative standard errors estimated for $F_{max}$ and $ED_{50}$ were 24.2% and 35.4%, respectively. Moreover, CsA bioavailability in patients receiving a PD of 80 mg was 20.6% lower than that in patients receiving a PD of 20 mg.

Multiple genetic variants in genes encoding the CsA-metabolising enzymes CYP3A4/5 and the multidrug resistance transporter ABCB1 were analysed [41]. However, no significant effects of the selected SNPs on CsA PK were found. Previous genotyping analysis results indicated that *ABCB1* C1236T and G2677T/A are in strong LD with C3435T. Thus, the influence of the *ABCB1* C1236T-G2677T/A-C3435T haplotypes on CsA PK was considered. The CL/*F* of the non-CGC haplotype carrier was 14.4% less than that of the CGC haplotype carrier 3 months following the operation (ΔOFV -15.0, $P < 0.001$).

In addition to pathophysiological and genetic factors, concomitant medications were also investigated, but none displayed significant relationships with PK parameters. After other covariates were tested, the incorporation of POD reduced the OFV by -79.1 ($P < 0.001$), indicating that other sources of time-related variability should be considered in further analyses.

In the final model, all retained covariates caused a significant increase in OFV upon removal. For these reasons, this model was accepted as the definitive final model. Eq. 3 shows the results of the covariate analysis for CsA CL:

$$CL = 32.6 \times (FFM/50)^{0.75} \times (HCT/30)^{-0.426}$$
$$\times (POD/30)^{0.0821} \times 0.856 \text{ (if POD > 90, non-CGC haplotype carriers)} \quad (3)$$

In this equation, influence scopes were adjusted according to their respective median values as determined from the dataset. The parameter estimates and associated precisions are shown in Table 3. The condition number of the final model was 240.1.

### 3.3 Model evaluation

The GOF plots for the base and final models are presented in Figure 2. Compared with the base model and the model incorporating only POD, the final model was greatly improved and showed no obvious bias. Over 99.5% (2515/2528) of the observations were within ±4 conditional weighted residuals.

The pcVPC results demonstrated good predictability of drug concentrations. The pcVPCs of the final model are depicted in Figure 3. The simulated data corresponded well with the observed data, indicating a lack of significant model misspecifications. Consistent with this, bootstrap parameter estimates closely matched the mean estimates from the population model, confirming model stability (Table 3).

### 3.4 Dosing regimen optimisation

The results of the Monte Carlo simulation are shown in Table S4. The predicted time course of CsA concentrations in a 'typical' patient (i.e., patients with median covariates), which was simulated based on 1,000 hypothetical individuals in different postoperative periods, is shown in Figure 4. A stable recommended dosing regimen could maintain median concentrations within a desired concentration range, indicating that the final model could be applied to design the dosing regimen.

## 4 Discussion

Although nearly 20 CsA popPK studies have been reported in adult renal transplant recipients, no large long-term follow-up cohort study has been conducted to elucidate the time-related variability in CsA PK [16]. Moreover, CsA PK characteristics may change with POD, thereby requiring dose adjustment to achieve target exposure. Therefore, it is essential to identify factors that can explain the variability in CsA PK and characterise the time-dependent CL/*F* of CsA.

To the best of our knowledge, this is the first popPK study of long-term changes in CsA CL/*F* with POD in adult renal transplant patients. A two-compartment model with first-order absorption and lag time adequately described the CsA PK properties. In the final model, FFM, HCT level, POD, and the presence of the *ABCB1* CGC haplotype were the most influential covariates with regard to CsA clearance. Tapering of the PD following the initial stage of transplantation may contribute to changes in bioavailability.

In a study conducted by Fanta *et al.*, as PK data on intravenously and orally administered CsA were available, potential factors affecting the absorption process, which may be related to long-term PK properties, were analysed [25]. Dose-dependent absorption or saturable presystemic metabolism contributed to this tendency. In this study, the factors that may influence CsA distribution were also considered, and we found that time-dependent clearance could be largely attributed to changes in erythrocyte binding owing to increased HCT levels with time after transplantation.

In the final model, the oral bioavailability of CsA increased by approximately 35.6% in the first month after transplantation, consistent with a previous report [25]. CsA exposure in the early phases after transplantation was 15.4% lower than that in the stable period when the same CsA dose was administered. The increased CsA bioavailability in the initial period may correlate with improved absorption in the intestine [42] and co-administration with steroids [43], which can activate the pregnane X receptor (PXR) to upregulate CYP3A and P-gp activity [35, 44]. Therefore, CsA dosage should be reduced when steroids are co-administered [20].

Approximately 58% of circulating CsA is bound to red blood cells [45]; therefore, changes in the HCT with POD may influence CsA PK. The HCT was low early in the postoperative period but increased during the first months after the transplantation, which is consistent with previous findings [46, 47]. The effect of the HCT on CsA clearance was retained in the final model, and a reduction in the HCT from 60.5% to 10.5% led to a 1.1-fold increase in CsA exposure. This relationship is consistent with known physiological properties [48]. Moreover, increases in the HCT led to the elevation of CsA binding to erythrocytes, which could partly prevent CsA extraction via the liver and distribution into peripheral tissues, resulting in elevated concentrations and decreased CL/*F* [21-23].

Multiple intrinsic/extrinsic factors, including demographic factors, genetic polymorphisms in drug-metabolising enzymes and transporters, disease progression, concomitant medications, and their combined effects, may influence the *in-vivo* behaviours of drugs in the clinic [49]. Most of these covariates contributed to *inter-* and *intra-*individual variability [16]. POD is a surrogate for many time-dependent factors [50]. Changes in covariates with POD were visually inspected and tested based on theoretical mechanisms in this study. More valuable information may be provided as these time-varying covariates can partly explain the effects of POD as a covariate on inter-individual variability [51]. Other currently unknown POD-related factors should be evaluated in future studies.

In addition to time-dependent factors, *ABCB1* genetic polymorphisms were also included in the final model. In this study, we found that *ABCB1* C1236T-G2677T/A-C3435T haplotypes may be an effective index for the characterisation of CsA PK. The CL/*F* of the non-CGC haplotype carrier was 14.4% lower than that of the CGC haplotype carrier 3 months post operation. P-gp is distributed widely on the brush border surface of the intestine and mainly influences the CsA absorption process [52]. Mutations at the above three sites may decrease the amount or activity of P-gp, resulting in the reduced excretion of CsA into the intestinal lumen and increased bioavailability [19, 53]. Fanta *et al*. reported that the *NR1I2* genotype, which encodes PXR, may influence CsA bioavailability [25]. However, no such information was available for this study, and additional studies are needed to evaluate this mechanism.

In this study, the recommended dosing regimens in different simulation scenarios were given based on the prior distributions of the final model. Clinicians can design an optimal regimen for each patient based on the individual's status. According to the simulation results in this study, different dose regimens were needed to achieve different target exposures during different postoperative periods. However, all of these regimens were obtained based on popPK parameter distributions in the final model. Enhanced individualised dose predictions can be designed via Bayesian forecasting using available concentrations as prior information [16, 54].

One potential limitation of this study was its retrospective observational design. Moreover, there was no way to confirm whether patients adhered to their prescribed dosage regimen. Additionally, this was a single-centre study. Multicentre validation is necessary to improve model predictability and assess the impact of CsA minimisation and precision-dosing strategies.

In conclusion, a popPK model for adult renal transplant recipients was developed based on a large long-term follow-up cohort study, and the time-dependent CL/*F* of CsA was comprehensively analysed. Allometrically scaled body size, HCT level, CGC haplotype carrier, and postoperative time may mediate CsA PK variability. Additionally, tapering of PD during the initial stage of transplantation may contribute to changes in bioavailability. Increases in HCT resulted in the enhancement of CsA binding to erythrocytes, leading to increased CsA concentrations and decreased CL/*F*, which may affect the time-dependent CL/*F* of CsA. Therefore, CsA dose adjustments should be considered during different postoperative periods.

# 5 Compliance with Ethical Standards

**Funding:**

This work was supported in part by grants from the '2019 Key Clinical Program of Clinical Pharmacy' (No. shslczdzk06502), and 'Weak Discipline Construction Project' (No. 2016ZB0301-01) of Shanghai Municipal Health and Family Planning Commission. We would like to thank Editage (www.editage.cn) for English language editing.

**Conflict of interest:**

There are no financial relationships with any organisations that might have an interest in the submitted work in the previous 3 years, and no other relationships or activities that could appear to have influenced the submitted work. The other authors have no conflicts of interest to declare.

**Ethical Approval:**

All procedures involving human participants were in accordance with the ethical standards of the institutional and/or national research committee and with the 1964 Helsinki Declaration and its later amendments or comparable ethical standards.

## 6 Contributors

JJM, XYQ and WWQ participated in the research design; JJM, XYQ, WWQ, LYX, MZ and MKZ implemented and conducted the study; JJM, XYQ and WWQ performed the research and analysed the data. JJM and WWQ drafted the manuscript, which was revised and approved by all the authors. There are no other relationships or activities that could appear to have influenced the submitted work.

## 7 Data availability statement

The data that support the findings of this study are available from the corresponding author upon reasonable request.

## Figure Legends

**Figure 1** Empirical Bayes estimates of cyclosporine clearance (CL/*F*) during the post-transplantation follow-up. The model-predicted typical CL/*F* (red dashed line) and the individual CL/*F* (blue circles) are shown.

**Figure 2** Diagnostic goodness-of-fit plots for base model, base model incorporating postoperative day (POD), and final model. (**a**) observations *vs.* population predictions; (**b**) observations *vs.* individual predictions; (**c**) conditional weighted residuals (CWRES) *vs.* population predictions; (**d**) CWRES *vs.* PODs. (**a** - **d**) The locally weighted regression line (red dashed lines). (**a**, **b**) the line of unity (black solid lines), and (**c**, **d**) y = 0 (solid lines) are shown.

**Figure 3** Prediction-corrected visual predictive checks (pcVPCs) stratified on postoperative days (PODs) for the final model, based on 2,000 simulations. The median observed values per bin (red solid line), the $5^{th}$ and $95^{th}$ percentiles (red dashed lines) of the observations, as well as the 95 % confidence interval of the $5^{th}$ and $95^{th}$ percentiles (blue areas) and the confidence interval of the median (red area) are shown.

**Figure 4** The concentration-time profiles of cyclosporine (CsA) from Monte-Carlo simulation of 1,000 hypothetical individuals with median covariates following different postoperative periods. The $25^{th}$ and $75^{th}$ percentiles of the simulation data (light blue), the median of the simulated data (red solid line), and both the therapeutic range of CsA predose concentrations (black solid lines) and 2-h postdose concentrations (black dashed lines) are shown. HCT, haematocrit; PD, prednisolone daily dose; POD, postoperative days. Simulation scheme with an asterisk (*) indicates *ABCB1* CGC haplotype carrier.

# Supporting information

Additional Supporting Information may be found in the online version of this article at the publisher's web-site:

*Supplementary Text S1* Target concentrations of CsA therapeutic monitoring

*Supplementary Text S2* Genotyping of *CYP3A4\*1G*, *CYP3A5\*3*, *ABCB1* C1236T, G2677T/A, and C3435T single-nucleotide polymorphisms

*Supplementary Text S3* The semi-mechanistic model used to predict fat-free mass

*Table S1* Characteristics of involved covariates in different postoperative periods

*Table S2* CsA sampling points distribution

*Table S3* Determination of ABCB1 C1236T-G2677T/A-C3435T haplotype with frequency and patient proportion more than 8%

*Table S4* Dose regimens recommended in different postoperative periods

*Figure S1* The tendency of covariates with postoperative days. The *red dashed line* represents the LOESS smoothing, and the *blue circles* represent the value of covariates.

**Table 1 Patients demographics used to develop and evaluate population model**

| Characteristics | Model development dataset | | Model evaluation dataset | |
|---|---|---|---|---|
| | Number or Mean ±SD | Median (Range) | Number or Mean ±SD | Median (Range) |
| No. of patients (Male/Female)[a] | 127 (81/46) | / | 56 (41/15) | / |
| No. of Samples ($C_0$/$C_2$/other)[b] | 2528 (1082/1180/266) | / | 1146 (541/605) | / |
| Age (years) | 40.4 ±10.1 | 40 (19-60) | 39.6 ±9.7 | 41 (18-58) |
| Height (cm) | 167.5 ±7.1 | 168.0 (150.0-188.0) | 168.5 ±9.5 | 170.0 (150.0-186.0) |
| Weight (kg) | 59.7 ±10.8 | 58.0 (40.0-95.0) | 62.0 ±11.0 | 61.0 (39.4-90.0) |
| Post-operation days | 426.0 ±877.8 | 24 (1-5998) | 443.1 ±695.0 | 111.5 (2-3942) |
| CsA daily dose (mg day$^{-1}$) | 278.6 ±92.0 | 300 (50-575) | 248.8 ±99.2 | 250.0 (50-600) |
| Prednisolone dose (mg day$^{-1}$) | 20.5 ±16.9 | 20 (0-80) | 12.0 ±15.1 | 7.5 (0-80) |
| $C_0$ (ng ml$^{-1}$) | 156.7 ±94.5 | 136.8 (22.6-974.6) | 149.4 ±100.7 | 123.6 (25.4-587.4) |
| $C_2$ (ng ml$^{-1}$) | 805.2 ±382.6 | 761.9 (108.8-2572.8) | 750.8 ±365.0 | 703.0 (34.6-2109.0) |
| Haematocrit (%) | 30.7 ±7.0 | 30.2 (10.5-60.5) | 34.7 ±7.9 | 34.7 (15.6-57.0) |
| Total Bilirubin (μmol L$^{-1}$) | 10.3 ±7.1 | 9.0 (1.0-168.9) | 12.1 ±6.4 | 10.8 (1.7-48.3) |
| Alanine aminotransferase (U L$^{-1}$) | 34.3 ±46.3 | 21.0 (4.0-420.0) | 39.8 ±35.1 | 19.0 (3.0-374.0) |
| Aspartate transferase (U L$^{-1}$) | 24.7 ±22.0 | 20.0 (5.0-383.0) | 22.5 ±18.2 | 18.0 (1.0-279.0) |
| Albumin (g L$^{-1}$) | 35.7 ±6.5 | 36.0 (20.0-52.0) | 37.6 ±6.5 | 37.4 (22.0-51.0) |
| Total protein (g L$^{-1}$) | 62.7 ±9.4 | 62.0 (41.0-88.0) | 66.7 ±8.9 | 67.0 (46.0-88.0) |
| Serum Creatinine (μmol L$^{-1}$) | 133.3 ±98.4 | 110.0 (14.0-1088.0) | 113.3 ±93.5 | 107.0 (48.0-776.0) |
| Creatinine Clearance (ml min$^{-1}$)[c] | 62.5 ±23.6 | 61.7 (6.2-360.7) | 67.2 ±24.7 | 67.1 (6.2-182.6) |
| Concomitant medications[a] | | | | |
|     Felodipine | 74 | / | 23 | / |
|     Nifedipine | 54 | / | 18 | / |
|     Perdipine | 14 | / | 7 | / |

$C_0$, pre-dose concentration; $C_2$, 2-hour post-dose concentration

[a] Data are expressed as number of patients

[b] Data are expressed as number of samples

[c] Calculated following the Cockcroft-Gault formula: CLcr = [(140-Age(year)) ×WT(kg)]/ (0.818×Scr (μmol L$^{-1}$)) ×(0.85 for female)

**Table 2 Allele frequencies of genetic polymorphisms in CYP3A4, CYP3A5 and ABCB1 genes**

| Single nucleotide polymorphisms | Number of recipients | Frequency (%) |
| --- | --- | --- |
| CYP3A4*1G (G82266A, rs 2242480) | | |
|   GG (*1/*1) | 95 | 56.9 |
|   GA (*1/*1G) | 63 | 37.7 |
|   AA (*1G/*1G) | 9 | 5.4 |
| CYP3A5*3 (A6986G, rs776746) | | |
|   AA (*1/*1) | 8 | 4.8 |
|   GA (*1/*3) | 75 | 44.9 |
|   GG (*3/*3) | 84 | 50.3 |
| ABCB1-C1236T (rs1128503) | | |
|   CC | 25 | 15.0 |
|   CT | 68 | 40.7 |
|   TT | 74 | 44.3 |
| ABCB1-G2677T/A (rs2032582) | | |
|   AA | 9 | 5.4 |
|   GG | 40 | 24.0 |
|   GA | 20 | 12.0 |
|   TT | 31 | 18.6 |
|   TG | 54 | 32.3 |
|   TA | 13 | 7.8 |
| ABCB1-C3435T (rs1045642) | | |
|   CC | 63 | 37.7 |
|   CT | 79 | 47.3 |
|   TT | 25 | 15.0 |

The allele frequencies are found to be in Hardy-Weinberg equilibrium ($P > 0.05$)

**Table 3 Parameter estimates for the base model with and without POD, the final model and the bootstrap procedure**

| Parameters | Base model | | Base model with POD | | Final model | | Bootstrap of final model | |
|---|---|---|---|---|---|---|---|---|
| | Estimate | RSE (%) | Estimate | RSE (%) | Estimate | RSE (%) | Median | 95% CI |
| Objective function value | 27908.6 | / | 27854.5 | / | 27737.5 | / | / | / |
| $K_a$ (h$^{-1}$) | 1.8 | 36.8 | 1.8 | 23.4 | 1.94 | 28.9 | 2.06 | 1.26 - 2.71 |
| CL/$F$ (L h$^{-1}$) | 30.8 | 3.1 | 29.5 | 4.9 | 32.6 | 5.0 | 32.2 | 28.2 - 35.2 |
| $V_c$/$F$ (L) | 115 | 9.9 | 117 | 14.7 | 127 | 13.0 | 128.5 | 103.5 - 152.4 |
| Q/$F$ (L h$^{-1}$) | 25.6 | 11.2 | 25.5 | 24.7 | 28 | 22.3 | 27.5 | 15.2 - 35.3 |
| $V_p$/$F$ (L) | 409 | 18.2 | 461 | 21.6 | 505 | 48.7 | 490.3 | 312.1 - 868.1 |
| $T_{lag}$ (h) | 0.454 | 5.7 | 0.454 | 5.6 | 0.453 | 4.3 | 0.459 | 0.407 - 0.486 |
| Covariate effect on CL/$F$ | | | | | | | | |
| HCT | / | / | / | / | -0.426 | 28.9 | -0.410 | -0.610 - (-0.209) |
| CGC[a] | / | / | / | / | 0.856 | 14.8 | 0.856 | 0.658 - 1.038 |
| POD | / | / | 0.00235 | 35.3 | 0.0821 | 46.5 | 0.0759 | 0.0348 - 0.142 |
| PD[b] | / | / | / | / | -0.288 | 24.2 | -0.301 | -0.656 - (-0.135) |
| | / | / | / | / | 23.8 | 35.4 | 29.6 | 4.0 - 95.7 |
| Between-subject variability | | | | | | | | |
| $K_a$ (%) | 90.4 | 22.6 | 89.7 | 17.5 | 88.5 | 20.6 | 94.8 | 69.2 - 129.7 |
| CL/F (%) | 24.2 | 9.6 | 25.1 | 10.0 | 25.4 | 14.6 | 25.0 | 20.2 - 31.1 |
| $V_c$/$F$ (%) | 31.8 | 16.1 | 31.0 | 20.8 | 26.8 | 15.3 | 25.4 | 15.5- 35.8 |
| Q/$F$ (%) | 33.5 | 40.9 | 40.0 | 17.6 | 36.3 | 26.6 | 37.2 | 11.7 - 55.4 |
| $V_p$/$F$ (%) | 129.2 | 13.7 | 142.1 | 20.9 | 135.3 | 26.9 | 134.4 | 94.1 - 211.9 |
| Residual variability | | | | | | | | |
| Proportional (%) | 37.9 | 4.0 | 37.3 | 3.7 | 36.7 | 3.6 | 36.3 | 33.7 - 38.7 |

CGC, *ABCB1* CGC haplotype carrier; CI, percentile confidence intervals; CL/$F$, apparent clearance; $F$, the bioavailability relative to 1; FFM, fat-free mass; HCT, haematocrit; $K_a$, absorption rate constant; PD, prednisolone daily dose; POD, postoperative days; Q/$F$, inter-compartmental clearance; $T_{lag}$, absorption lag time; $V_c$/$F$, apparent central volume of distribution; $V_p$/$F$, apparent peripheral volume of distribution

[a] the CL/$F$ of non-CGC haplotype carriers was 0.856 times to CGC haplotype carriers since postoperative three months

[b] -0.288 is the maximum change in $F$ with increasing prednisolone daily dose; 23.8mg is the prednisolone daily dose with half maximum on $F$

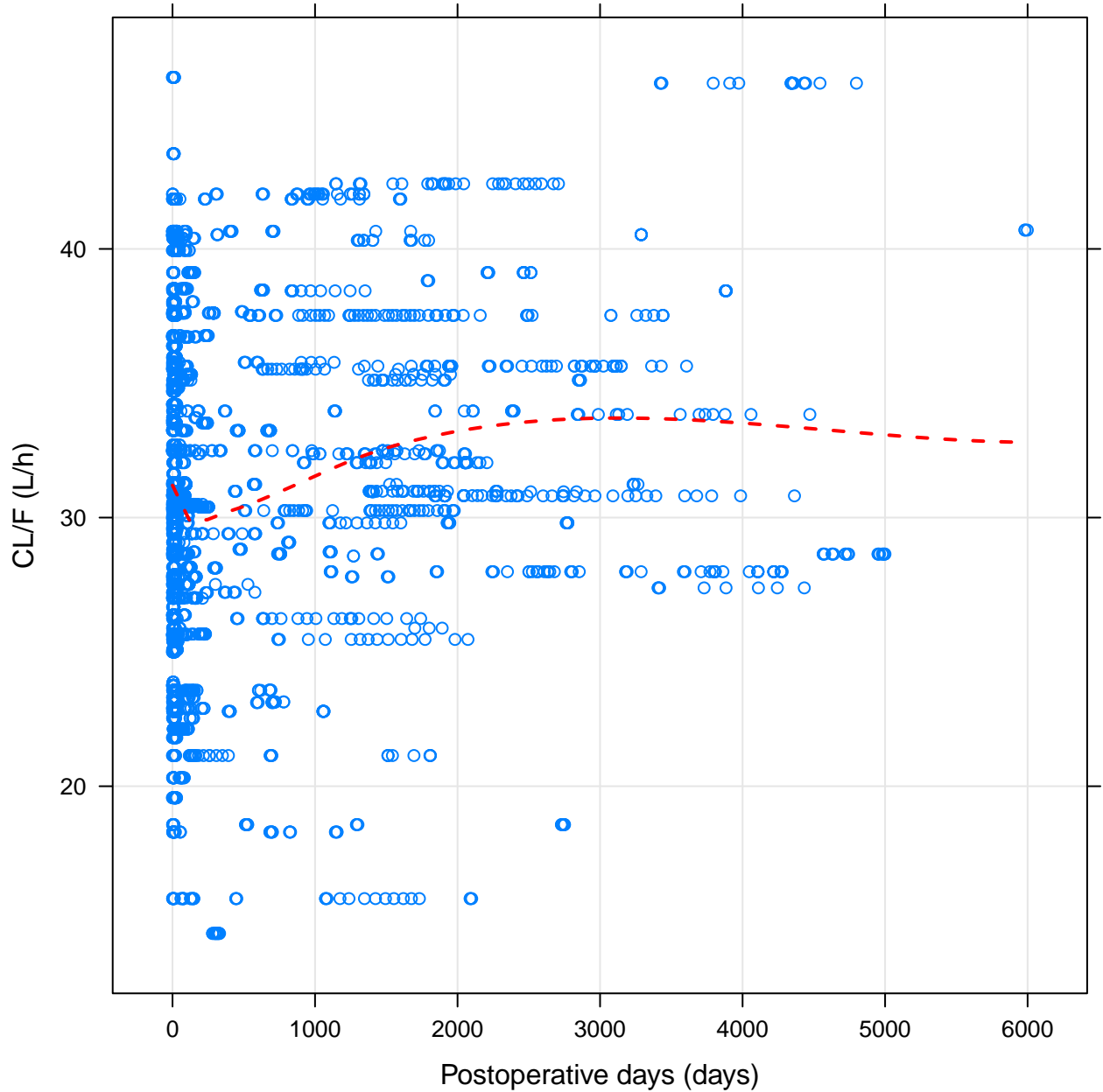

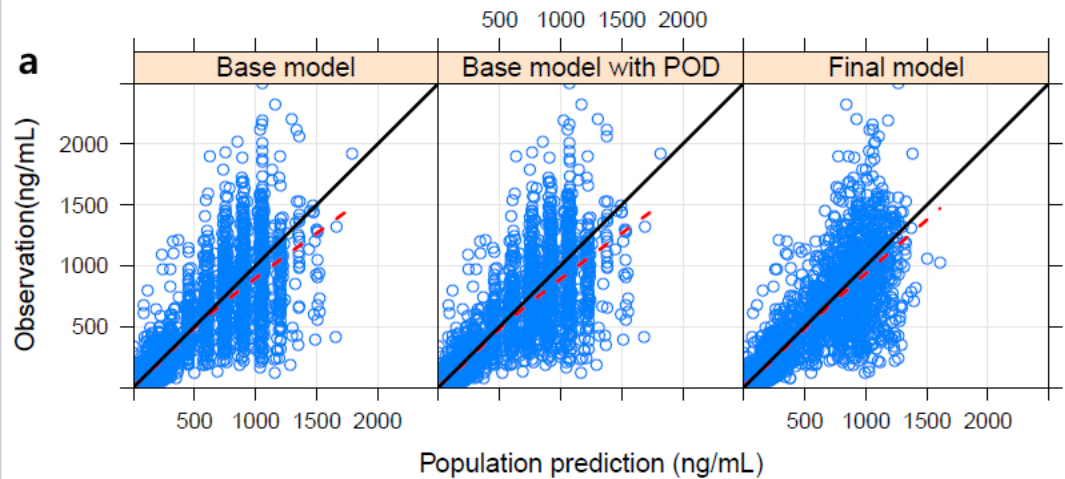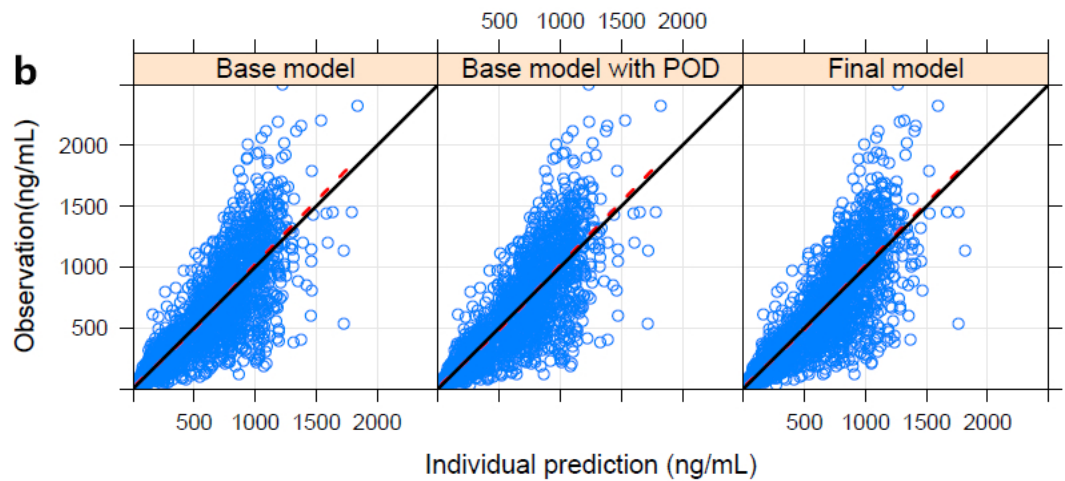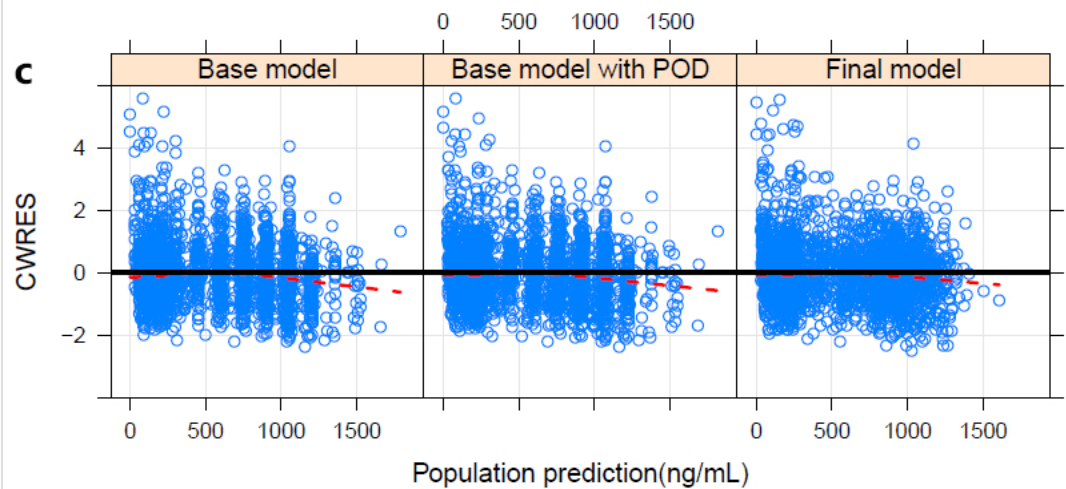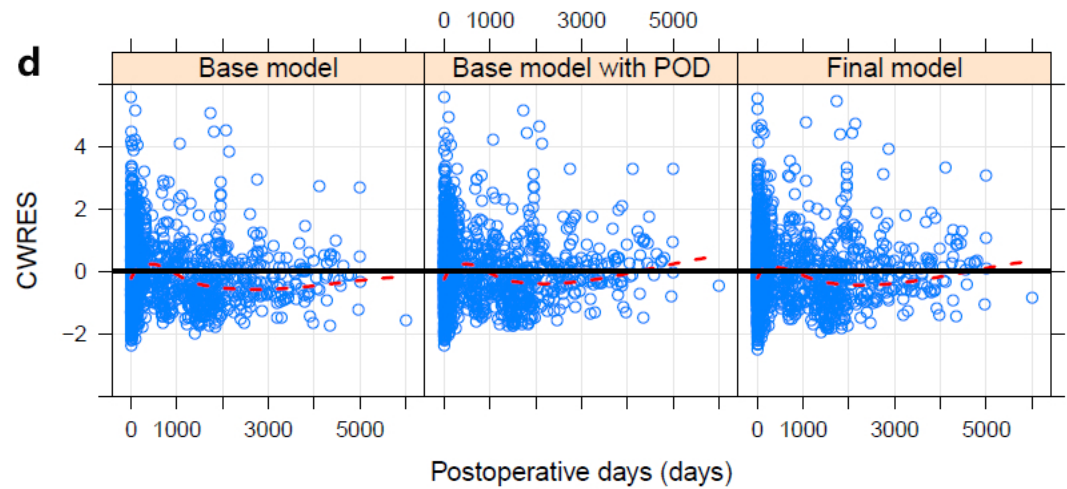

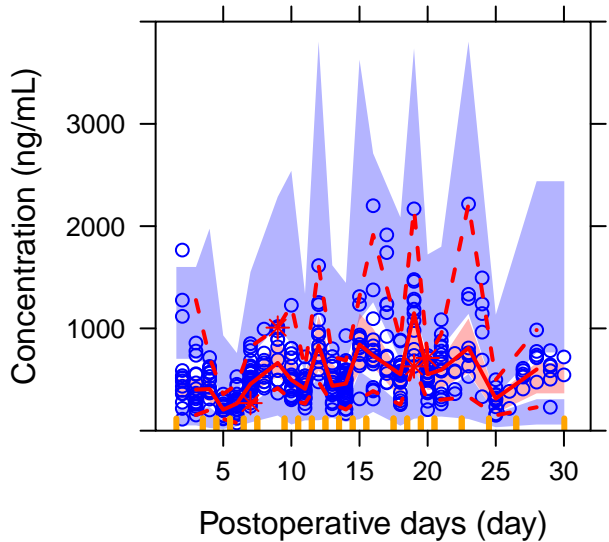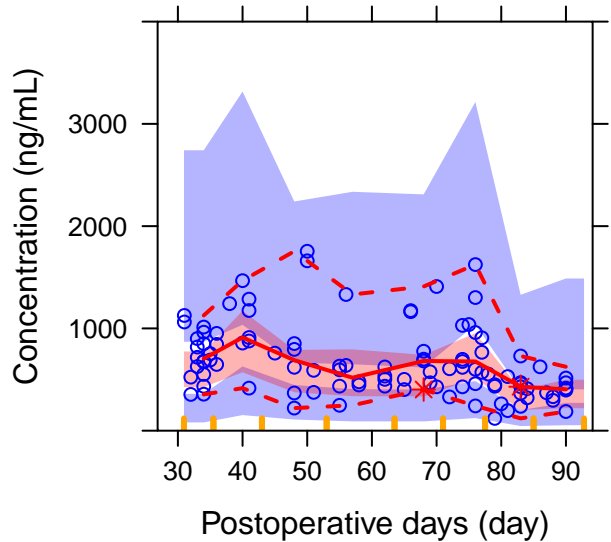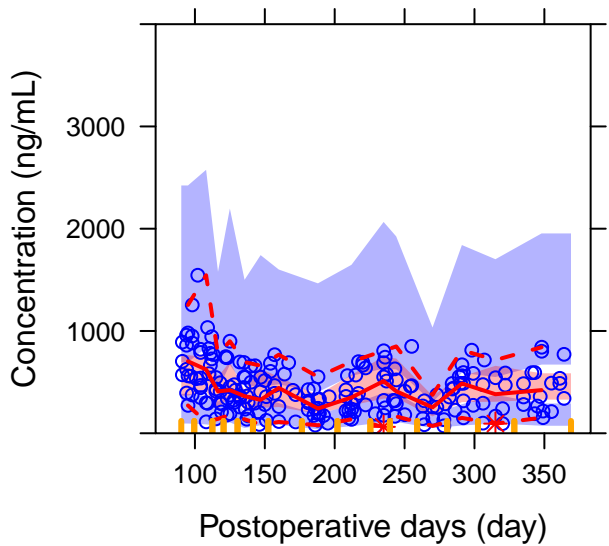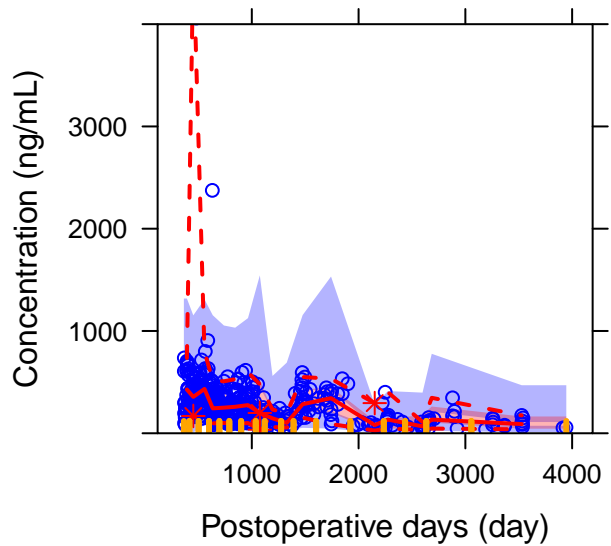

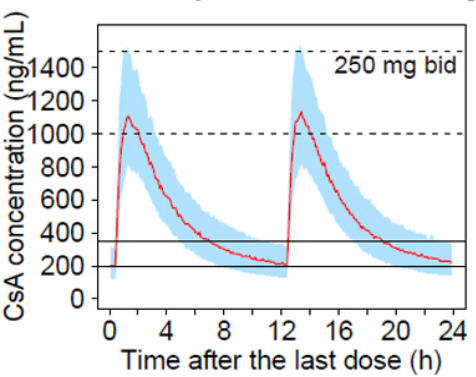
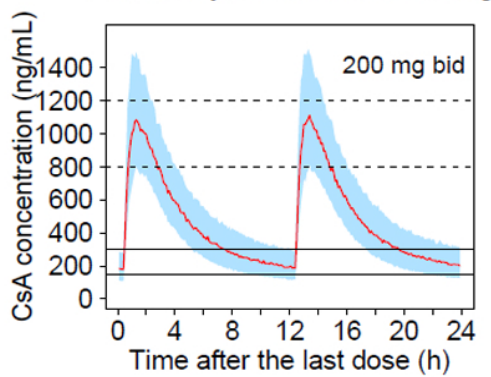
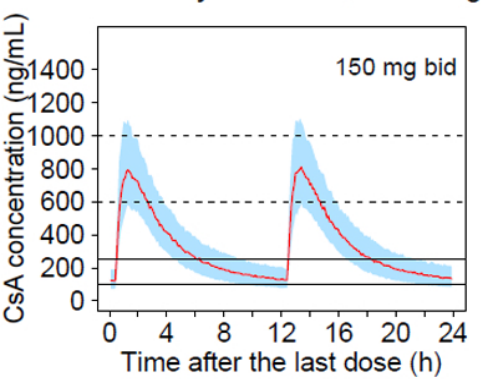
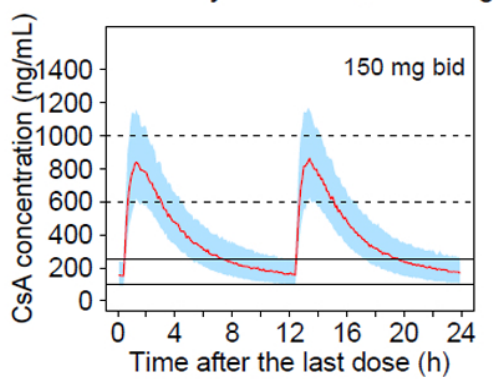
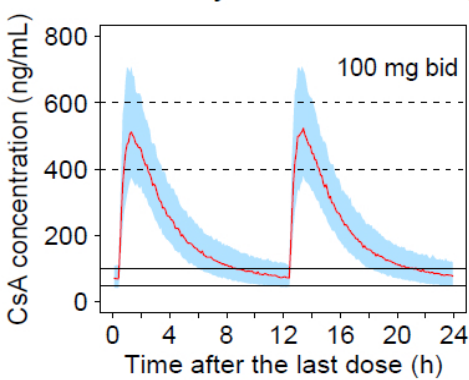
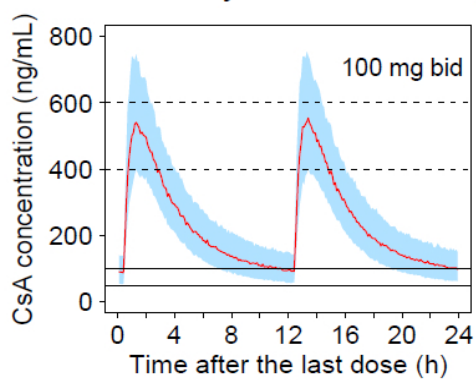
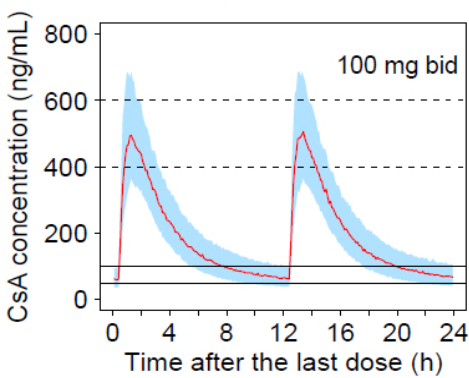
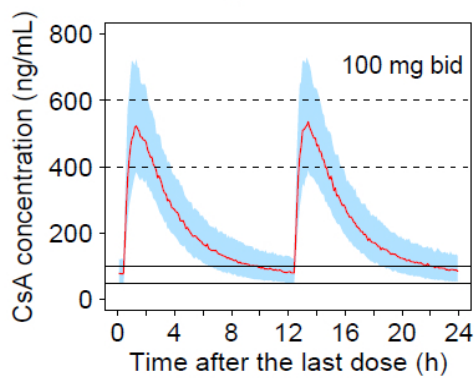

# Electronic Supplementary Material

**Supplementary Text S1 Target concentrations of CsA therapeutic monitoring**

Target $C_0$ values were 200 - 350 ng ml$^{-1}$ in the first month, 150 - 300 ng mL$^{-1}$ during months 1 to 3, 100 - 250 ng mL$^{-1}$ during months 3 to 12, and 50 - 100 ng mL$^{-1}$ thereafter; target $C_2$ values were 1000 - 1500 ng mL$^{-1}$ in the first month, 800 - 1200 ng mL$^{-1}$ during months 1 to 3, 600 - 1000 ng mL$^{-1}$ during months 3 to 12, and 400 - 600 ng mL$^{-1}$ thereafter.

**Supplementary Text S2 Genotyping of CYP3A4*1G, CYP3A5*3, ABCB1 C1236T, G2677T/A, and C3435T single-nucleotide polymorphisms**

Genomic DNA was extracted from whole blood using TIANamp Blood DNA Kit (TIANGEN, DP318). Polymerase chain reaction (PCR) was performed in a reaction volume of 20μL, containing 10μL MIX, 7μLdd$H_2O$, 1μL forward primer, 1μL backward primer and 1μL DNA template. The sequences of forward and reverse primers were listed below.

**The sequences of primers for genotyping SNPs of CYP3A5 and ABCB1**

| SNP | Primer |
|---|---|
| CYP3A4*1G (rs2242480) | F: 5'CACCCTGATGTCCAGCAGAAACT3' |
| | R: 5'AATAGAAAGCAGATGAACCAGAGCC3' |
| CYP3A5*3 (rs776746) | Forward: 5'CTTTAAAGAGCTCTTTTGTCTCTC3' |
| | Reverse: 5'CCACGAAGCCAGACTTTGAT3' |
| ABCB1 C1236T (rs1128503) | Forward: 5'TCTTTGTCACTTTATCCAGC3' |
| | Reverse: 5'TCTCACCATCCCCTCTGT3' |
| ABCB1 G2677T/A (rs2032582) | Forward: 5'TGCAGGCTATAGGTTCCAGG3' |
| | Reverse: 5'TTTAGTTTGACTCACCTTCCCG3' |
| ABCB1 C3435T (rs1045642) | Forward: 5'TGCTGGTCCTGAAGTTGATCTGTGAAC3' |
| | Reverse:5'ACATTAGGCAGTGACTCGATGAAGGCA3' |

Cycle conditions were: initial denaturation at 95 ℃ for 5 min, then 35 cycles of denaturation at 95 ℃ for 15s, annealing at 55 ℃ for 15s and elongation at 72 ℃ for 30s, followed by a final extension at 72 ℃ for 5 min. PCR products were analyzed by electrophoresis with 2% agarose gels (DNA Marker DL2000) and purified using PCR Product Purification Kit (Bodataike Bioengineering Corporation, Peiking, China).

**Supplementary Text S3 The semi-mechanistic model used to predict fat-free mass (FFM)**

The semi-mechanistic model be used to predict FFM as follows [1]:

$$\text{FFM (male)} = \frac{9270 \times \text{WT}}{6680 + 216 \times \text{BMI}} \qquad (1)$$

$$\text{FFM (female)} = \frac{9270 \times \text{WT}}{8780 + 244 \times \text{BMI}} \qquad (2)$$

Where WT is measured in kilograms, BMI is calculated following the formula: BMI = weight (kg)/height$^2$ (m$^2$), *BMI* body mass index, *WT* body weight

**References**

1. Janmahasatian S, Duffull SB, Ash S, Ward LC, Byrne NM, Green B. Quantification of lean bodyweight. Clinical pharmacokinetics 2005; 44: 1051-65.

**Supplementary Table S1 Characteristics of involved covariates in different postoperative periods**

|  | < 1Month | 1-3 Months | 3-12 Months | > 12 Months |
|---|---|---|---|---|
| Fat-free mass (kg) | 47.6 (30.1-58.9) | 49.2 (31.4-58.8) | 51.3 (33.2-63.1) | 49.8 (32.8-61.7) |
| Hematocrit (%) | 27.3 (19.3-35.3) | 34.8 (25.0-42.6) | 36.5 (24.6-46.5) | 38.4 (25.0-47.8) |
| Prednisolone dose (mg day$^{-1}$) | 20.0 (0-70.0) | 20.0 (10.0-20.0) | 10.0 (5.0-20.0) | 5.0 (0-12.4) |

Data are expressed as median (5% - 95% percentiles)

Patients with median covariates were considered as standard

**Supplementary Table S2 CsA sampling points distribution**

| Sampling schedule | Number of samples | Frequency (%) |
|---|---|---|
| predose | 1623 | 44.18 |
| 0.5 hour | 24 | 0.65 |
| 1 hour | 28 | 0.76 |
| 1.5 hour | 24 | 0.65 |
| 2 hour | 1785 | 48.58 |
| 3 hour | 57 | 1.55 |
| 4 hour | 39 | 1.06 |
| 6 hour | 26 | 0.71 |
| 8 hour | 24 | 0.65 |
| 10 hour | 20 | 0.54 |
| 12 hour | 24 | 0.65 |
| Total | 3674 | 100 |

**Supplementary Table S3 Determination of ABCB1 C1236T-G2677T/A-C3435T haplotype with frequency and patient proportion more than 8%**

| Haplotype | C1236T | G2677T/A | C3435T | Number | Total number | Proportion (%) |
|---|---|---|---|---|---|---|
| TTT | TT | TT | TT | 18 | 31 | 18.56 |
|  | TT | TT | CT | 9 |  |  |
|  | TT | GT | TT | 2 |  |  |
|  | TT | TA | TT | 0 |  |  |
|  | CT | TT | TT | 2 |  |  |
| TGC | TT | GG | CC | 11 | 28 | 16.77 |
|  | TT | GG | CT | 3 |  |  |
|  | CT | GG | CC | 8 |  |  |
|  | TT | GT | CC | 6 |  |  |
|  | TT | GA | CC | 0 |  |  |
| CGC | CC | GA | CC | 8 | 25 | 14.97 |
|  | CC | GG | CC | 7 |  |  |
|  | CC | GG | CT | 2 |  |  |
|  | CT | GG | CC | 8 |  |  |
|  | CC | GT | CC | 0 |  |  |
| CAC | CC | AA | CC | 6 | 16 | 9.58 |
|  | CC | AA | CT | 1 |  |  |
|  | CT | AA | CC | 1 |  |  |
|  | CC | GA | CC | 8 |  |  |
|  | CC | TA | CC | 0 |  |  |

**Supplementary Table S4 Dose regimens recommended in different postoperative periods**

| Simulation scheme | FFM (kg) | PD (mg) | HCT (%) | POD (days) | CGC haplotype carrier status | Dose regimen recommended | $C_0$[a] | $C_2$[a] |
|---|---|---|---|---|---|---|---|---|
| 01a | 50 | 80 | 19.3 | 4 | / | 300 mg bid | 200.4 (124.0-322.5) | 1167.7 (855.3-1531.8) |
| 01b | 50 | 80 | 27.3 | 4 | / | 250 mg bid | 210.3 (129.0-335.7) | 1046.3 (769.1-1377.6) |
| 01c | 50 | 80 | 35.3 | 4 | / | 250 mg bid | 245.0 (153.4-393.1) | 1105.0 (818.2-1445.6) |
| 02a | 50 | 20 | 25.0 | 30 | / | 200 mg bid | 152.2 (93.7-245.1) | 952.9 (699.2-1245.7) |
| 02b | 50 | 20 | 34.8 | 30 | / | 200 mg bid | 190.8 (116.7-308.2) | 1021.3 (749.9-1342.1) |
| 02c | 50 | 20 | 42.6 | 30 | / | 175 mg bid | 189.3 (117.4-303.6) | 932.0 (686.4-1227.6) |
| 03a | 50 | 10 | 24.6 | 90 | carrier | 175 mg bid | 114.6 (69.8-184.6) | 790.6 (581.9-1042.3) |
| 03b | 50 | 10 | 36.5 | 90 | carrier | 150 mg bid | 128.2 (79.1-206.3) | 740.6 (542.5-971.8） |
| 03c | 50 | 10 | 46.5 | 90 | carrier | 150 mg bid | 150.7 (92.2-241.2) | 777.1 (571.3-1023.6) |
| 04a | 50 | 10 | 24.6 | 90 | non-carrier | 150 mg bid | 125.0 (77.1-201.1) | 733.9 (537.9-963.4) |
| 04b | 50 | 10 | 36.5 | 90 | non-carrier | 150 mg bid | 161.2 (99.8-258.5) | 796.8 (587.2-1050.3) |
| 04c | 50 | 10 | 46.5 | 90 | non-carrier | 150 mg bid | 185.8 (117.2-299.1) | 837.6 (620.3-1097.4) |
| 05a | 50 | 5 | 25.0 | 365 | carrier | 100 mg bid | 54.5 (32.6-88.7) | 428.4 (315.2-562.5) |
| 05b | 50 | 5 | 38.4 | 365 | carrier | 100 mg bid | 73.4 (45.1-118.0) | 470.5 (345.0-616.1) |
| 05c | 50 | 5 | 47.8 | 365 | carrier | 100 mg bid | 85.3 (52.6-137.3) | 493.5 (361.5-647.6) |
| 06a | 50 | 5 | 25.0 | 365 | non-carrier | 100 mg bid | 70.6 (43.2-112.6) | 463.3 (339.7-608.4) |
| 06b | 50 | 5 | 38.4 | 365 | non-carrier | 100 mg bid | 94.2 (57.7-151.7) | 508.6 (373.1-668.7) |
| 06c | 50 | 5 | 47.8 | 365 | non-carrier | 75 mg bid | 80.9 (50.1-129.7) | 399.0 (294.0-525.7) |
| 07a | 50 | 0 | 25.0 | 1095 | carrier | 125 mg bid | 58.5 (34.9-96.1) | 510.2 (373.9-667.4) |
| 07b | 50 | 0 | 38.4 | 1095 | carrier | 100 mg bid | 63.8 (38.8-103.1) | 448.0 (330.4-591.0) |
| 07c | 50 | 0 | 47.8 | 1095 | carrier | 100 mg bid | 73.8 (45.4-118.7) | 471.5 (345.8-617.2) |
| 08a | 50 | 0 | 25.0 | 1095 | non-carrier | 100 mg bid | 61.0 (37.0-98.9) | 442.1 (326.5-582.1) |
| 08b | 50 | 0 | 38.4 | 1095 | non-carrier | 100 mg bid | 81.7 (50.5-131.6) | 486.5 (356.3-637.4) |
| 08c | 50 | 0 | 47.8 | 1095 | non-carrier | 100 mg bid | 94.7 (57.9-152.8) | 509.5 (373.8-669.6) |

$C_0$, pre-dose concentration; $C_2$, 2-hour post-dose concentration; CGC, *ABCB1* CGC haplotype carrier; FFM, fat-free mass; HCT, haematocrit; PD, prednisolone daily dose; POD, postoperative days

[a] Data are expressed as median (25% - 75% percentiles)

# Electronic Supplementary Material

**Supplementary Fig. S1**

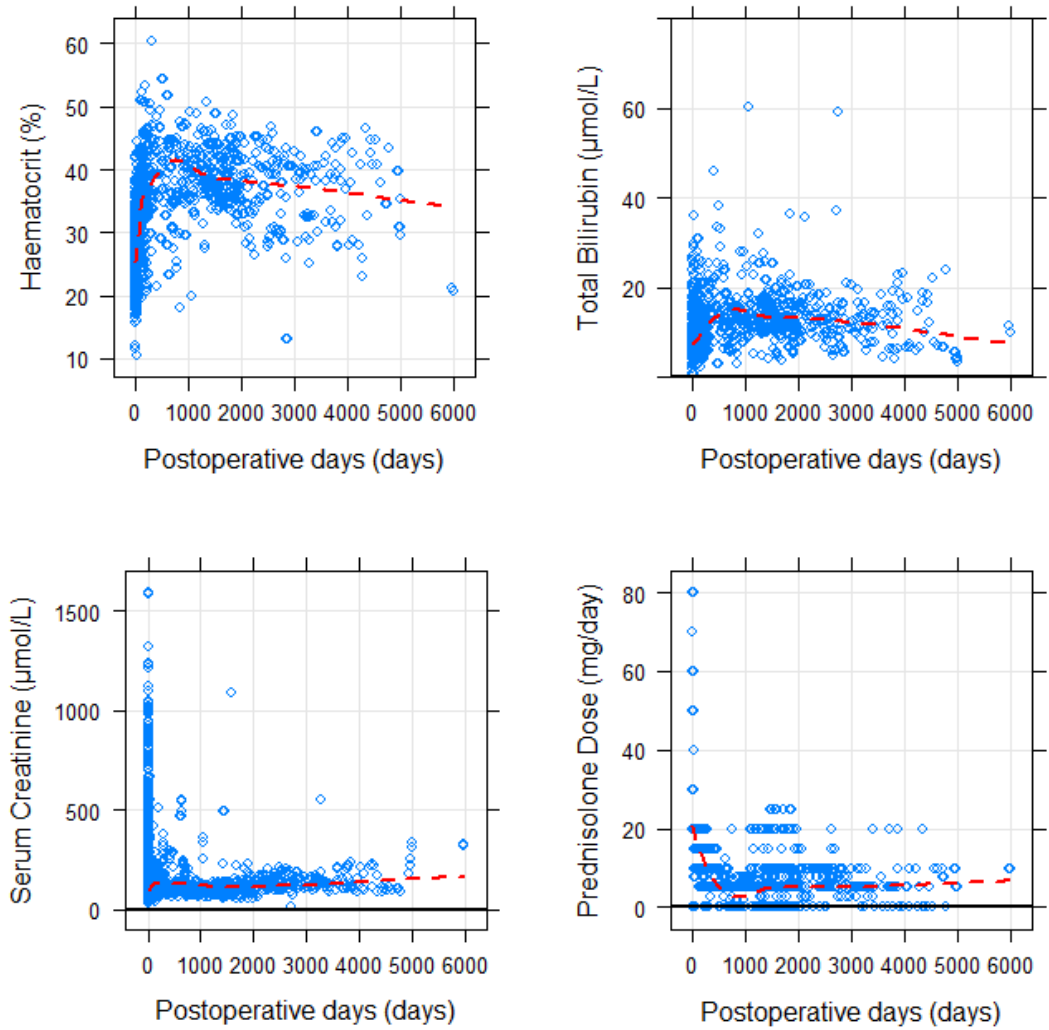